\documentclass[aps]{revtex4}

\usepackage{amsmath, amssymb, amsthm, amsfonts, latexsym, graphicx}

\newtheorem*{lem1}{Lemma 1}
\newtheorem*{lem2}{Lemma 2}
\newtheorem*{lem3}{Lemma 3}

\newcommand{\ket}[1]{\left| #1\right\rangle}      
\newcommand{\bra}[1]{\left\langle #1\right|}      
\newcommand{\kets}[1]{| #1 \rangle}    				    
\newcommand{\ii}{\mathbb{I}}											

\newcommand{\half}{\frac{1}{2}}

\newcommand{\cH}{{\mathcal H}} 
\newcommand{\tav}[1]{       
	\frac{1}{\tau_{10}}\int^{\tau_{10}}_0 
	#1 \,d\tau
}
\newcommand{\tur}{\,\circlearrowleft\,}   
\newcommand{\gat}{\,\blacktriangleright}  
\newcommand{\mov}{\,\vartriangleright}    
\newcommand{\bul}{\:\:\centerdot\:}       
\newcommand{\iga}{\:I\,}                  
\newcommand{\ici}{{\bigcirc \mkern-13mu   
	\mbox{\scriptsize{\textit{I}}}\mkern5mu\,}}
\newcommand{\wga}{W}						
\newcommand{\wci}{{\bigcirc \mkern-17mu 	
	\mbox{\scriptsize{\textit{W}}}\mkern1mu\,}}
\newcommand{\sga}{\:S\,}					
\newcommand{\sci}{{\bigcirc \mkern-14mu 	
	\mbox{\scriptsize{\textit{S}}}\mkern3mu\,}}
\newcommand{\aga}{\,A\,}					
\newcommand{\aci}{{\bigcirc \mkern-14mu 	
	\mbox{\scriptsize{\textit{A}}}\mkern1mu\,}}
\newcommand{\bga}{\,B\,}					
\newcommand{\bci}{{\bigcirc \mkern-15mu 	
	\mbox{\scriptsize{\textit{B}}}\mkern4mu\,}}
\newcommand{\goes}{\:\rightarrow\:}		
\newcommand{\band}[2]{		
	\begin{array}{|r|r|}
	\hline #1 & #2 \\
	\hline
	\end{array}}
\newcommand{\triUR}[3]{ 		
	\begin{array}{r@{}c|c|}
	\cline{2-3} \vline & \:\: #1 & #2 \\ 
	\cline{2-3} & & #3 \\ 
	\cline{3-3} 
	\end{array}\:\:}
\newcommand{\triUL}[3]{ 		
	\:\:\begin{array}{|c|c@{}l}
	\cline{1-2} #1 & #2 \:\: & \vline \\ 
	\cline{1-2} #3 & & \\ 
	\cline{1-1}
	\end{array}}
\newcommand{\four}[4]{ 		
	\begin{array}{|c|c|}
	\hline #1 & #2 \\
	\hline #3 & #4 \\
	\hline
	\end{array}}
\newcommand{\flour}[3]{ 		
	\begin{array}{|c|c|}
	\hline #1 & #2 \\
	\hline \multicolumn{2}{|c|}{#3} \\
	\hline
	\end{array}}

\newlength{\onebox}
\newcommand\raiseonebox{\raisebox{-.5\onebox} 
  {\rule{0pt}{\onebox}}}

\begin{document}

\title{Hamiltonian Quantum Cellular Automata in 1D}

\author{Daniel Nagaj}
\affiliation{Center for Theoretical Physics, MIT, Cambridge, MA 02139, USA}
\affiliation{Research Center for Quantum Information, Slovak Academy of Sciences, D´ubravsk´a cesta 9, 845 11 Bratislava, Slovakia}
\affiliation{Quniverse, L´i¡s¡cie ´udolie 116, 841 04, Bratislava, Slovakia}
\email{daniel.nagaj@savba.sk}
\author{Pawel Wocjan}
\affiliation{School of Electrical Engineering and Computer Science, University of Central Florida, Orlando, FL 32816, USA}
\email{wocjan@cs.ucf.edu}

\date{April 22, 2008}

\begin{abstract}
We construct a simple translationally invariant, nearest-neighbor Hamiltonian on a chain of $10$-dimensional qudits that makes it possible to realize universal quantum computing without any external control during the computational process. We only require the ability to prepare an initial computational basis state which encodes both the quantum circuit and its input. The computational process is then carried out by the autonomous Hamiltonian time evolution. 
After a time polynomially long in the size of the quantum circuit has passed, the result of the computation is obtained with high probability by measuring a few qudits in the computational basis.

This result also implies that there cannot exist efficient classical simulation methods for generic translationally invariant nearest-neighbor Hamiltonians on qudit chains, unless quantum computers can be efficiently simulated by classical computers (or, put in complexity theoretic terms, unless BPP=BQP).
\end{abstract}

\maketitle


\section{Introduction}

One of the most important challenges in quantum information science is to identify quantum systems that can be controlled in such a way that they can be used to realize universal quantum computing. The quantum circuit model abstracts from the details of concrete physical systems and states that the required elementary control operations are: (i) initialization in basis states, (ii) implementation of one and two-qubit gates, and (iii) measurement of single qubits in basis states.  Meanwhile, many other models have been proposed such as measurement-based quantum computing \cite{RB-oneway:00,Nielsen-Measurement:03,Leung-Measurement:04,CLN-Measurement:05}, adiabatic quantum computing \cite{Farhi-Adiabatic:01,Aharonov-Adiabatic:04}, or topological quantum computing \cite{Kitaev:03} that reduce or modify the set of elementary control operations.  However, the common principle underlying all these models is that the computation process is always driven by applying a sequence of control operations.

Instead, we consider a model that does not require any control during the computational process. This model consists of a quantum system with a Hamiltonian that makes it possible to realize universal quantum computing by the following protocol: (1) prepare an initial state in the computational basis that encodes both the program and input, (2) let the Hamiltonian time evolution act undisturbed for a sufficiently long time, and (3) measure a small subsystem in the computational basis to obtain the result of the computation with high probability.  We refer to this model as a {\em Hamiltonian quantum computer} and more specifically as a {\em Hamiltonian quantum cellular automaton} (HQCA) provided that the Hamiltonian acts on qudits that are arranged on some lattice, is invariant with respect to translations along the symmetry axis of the lattice, and contains only finite range interactions. Most natural Hamiltonians have these properties, so it is important to construct HQCA that are as close as possible to natural interactions.

Hamiltonian QCA are related to the more usual discrete-time QCA (for further review of the different types of quantum cellular automata we refer the reader to \cite{PDC-ModelsQCA:05}). However, while the evolution of discrete-time QCA proceeds in discrete update steps (corresponding to tensor products of local unitary operations, see e.g. \cite{Raussendorf-QCA:05,ShepherdPRL:06}), the states of Hamiltonian QCA change in a continuous way according to the Schr\"odinger equation (with a time-independent Hamiltonian).
For this reason, Hamiltonian QCA are also called {\em continuous-time} QCA \cite{PDC-ModelsQCA:05}. 
Also, in the HQCA model, all the couplings (interactions)
are present all the time, while for the the discrete-time QCA,
the execution of updates on overlapping cells is synchronized by external control.
Therefore, the nearest-neighbor interactions of a HQCA have to include a mechanism that ensures that the logical transformations are carried out in the correct order. 

The motivation to consider Hamiltonian computers is threefold.  First, it is a fundamental question in the thermodynamics of computation how to realize computational processes within a closed physical system.  Such Hamiltonian computers were presented and discussed by Benioff \cite{Benioff:80}, Feynman \cite{Feynman:85}, and Margolus \cite{Margolus:90}. Second, Hamiltonian quantum cellular automata could lead to new ideas for reducing the set of necessary control operations in current proposals for quantum computing by using the inherent computational power of the interactions.  HQCA are at one end of the spectrum of possible implementations; more realistic perspectives for quantum computing could arise by combining this model with more conventional models involving external control operations throughout the computation. Third, this model can show the limitations of current and future methods in condensed matter physics for simulating the time evolution of translationally invariant systems. If evolving with a certain Hamiltonian can realize universal quantum computing, then there cannot exist any classical method for efficiently simulating the corresponding time evolution unless classical computers are as powerful as quantum computers (BPP=BQP).

The first theoretical computational models based on a single time-independent Hamiltonian go back to \cite{Benioff:80, Feynman:85, Margolus:90}. However, these Hamiltonian computers were not explicitly designed for realizing universal {\em quantum} computing. Margolus' model \cite{Margolus:90} has the attractive feature that it is laid out on a
$2$-dimensional lattice with translationally invariant, finite-range interactions. (In \cite{Biafore:94} it was argued that the part of the Hamiltonian responsible for the synchronization in a $1$-dimensional variant is close to real interaction in solid states.) However,  this scheme does not satisfy the requirement (1) since its initial state has to be prepared in a superposition. Building upon Margolus' idea, a translationally invariant Hamiltonian 
universal for quantum computing even if the initial state is restricted to be a canonical basis state was given in \cite{JW:05}. This model requires $10$-local, finite-range interactions among qubits on a $2$-dimensional rectangular lattice wrapped around a cylinder.
Subsequently, it was established in \cite{Janzing:07} that nearest-neighbor interactions among qutrits on a $2$-dimensional lattice suffice. However, the Hamiltonian of \cite{Janzing:07} is translationally invariant only when translated over several lattice sites.
A different approach was taken by Vollbrecht and Cirac in \cite{VC-QCA:07}, showing that one can implement universal quantum computation with a translationally invariant, nearest-neighbor Hamiltonian on a chain of $30$-dimensional qudits. 
Also, recently another 1D translationally invariant Hamiltonian computer construction was given by Kay \cite{Kay31}, using particles with dimension $d=31$. 

We present two different simplified HQCA constructions on one-dimensional qudit chains. In both models, we think of the qudit chain as composed of two registers, data and program. The work qubits we compute on are located at a static location in the data register. Driven by the autonomous Hamiltonian time evolution, the program sequence contained in the program register moves past the work qubits and the gates are applied to them. After we let the system evolve for a time not larger than a polynomial 
in the length of the program, we measure one or two qudits in the computational basis to read out the output of the computation with high probability.

Our first construction is for a chain of $10$-dimensional qudits and is related to the ideas of \cite{VC-QCA:07}. The mechanism behind the progress of the program sequence in this particular model can be thought of as the diffusion of a system of free fermions on a line. Concurrently with our preprint, Chase and Landahl \cite{ChaseLandahl} found another 1D construction with particles with dimension $d=8$. However, their Hamiltonian is not translationally invariant. If we release the translational invariance requirement in our $d=10$ model, we obtain a $d=8$ construction as well. Whether the required dimension $d=10$ can be decreased while keeping translational invariance in our HQCA model remains an open question.

Our second construction uses qudits with dimension $d=20$ and is inspired by \cite{JWZ-measuring:07}, utilizing a technique of \cite{GottesmanLine:07} to transport the program. Here, the mechanism for the progress of the computation can be thought of as a quantum walk on a line.

The paper is organized as follows. First, in Section \ref{d10construct} we present the HQCA construction with cell size $d=10$
and analyze the required run-time of this model in Section \ref{d10time}. Second, we give the HQCA construction with cell size $d=20$ in Section \ref{d20construct} and discuss the readout procedure and the required run-time in Section \ref{d20time}.
We provide some useful results for the continuous time quantum walk on a line in Appendix \ref{d20proof} and prove a lemma concerning the diffusion of free fermions on a line in Appendix \ref{d10proof}.


\section{The $d=10$ Hamiltonian Quantum Cellular Automaton}
\label{d10section}

We present a simple universal HQCA on a chain of qudits with dimension $d=10$. First, we encode the progression of a quantum circuit $U$ on $N$ qubits into a set of states $\kets{\varphi_{\sigma}}$ of a chain of qudits with length $L=poly(N)$. Second, we give a translationally invariant nearest-neighbor Hamiltonian on this chain of qudits, which induces a quantum walk on the set of states $\kets{\varphi_{\sigma}}$. Finally, using a mapping to a system of free fermions in 1D, we prove that when we initialize the qudit chain in an easily determined computational basis state and let the system evolve for a time $\tau\leq \tau_{10} = O(L\log L)$ chosen uniformly at random, we can read out the result of the quantum circuit $U$ with probability
$p_{10}\geq \frac{5}{6} - O\left(\frac{1}{\log L}\right)$ by measuring one of the qudits in the computational basis. We then show that this is enough to ensure universality of our HQCA for the class BQP.


\subsection{The Construction}
\label{d10construct}

\subsubsection{Encoding a quantum circuit}

\begin{figure}
	\begin{center}
	\includegraphics[width=4in]{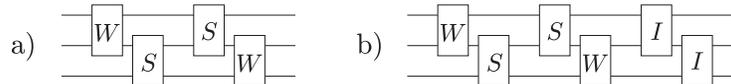}
	\caption{a) A quantum circuit consisting of two rounds of gates acting on nearest neighbors. b) The previous circuit with a third round of identity gates added.}
	\label{CircuitFigure}
	\end{center}
\end{figure}

The gate set \{Toffoli, Hadamard\} is universal for quantum computation \cite{Shi:02}. With only polynomial overhead, one can simulate a circuit consisting of these gates using only the gate $W$ (controlled $\frac{\pi}{2}$ rotation about the $y$-axis) 
\begin{eqnarray}
	W = \left[
		\begin{array}{cccc}
			1 & 0 & 0 & 0\\
			0 & 1 & 0 & 0 \\
			0 & 0 & \frac{1}{\sqrt{2}} & -\frac{1}{\sqrt{2}} \\
			0 & 0 & \frac{1}{\sqrt{2}} & \frac{1}{\sqrt{2}}
		\end{array}
	\right]
\end{eqnarray}
if it can be applied to any pair of qubits.
Let us consider implementing universal quantum computation on a qubit chain using only nearest neighbor gates. Let us also restrict the use of the $W$ gate so that the control qubit has to be to the left of the target qubit. Using only polynomially many additional swap gates $S$, one can still do universal quantum computation on a qubit chain. 
Thus given a quantum circuit $U'$ on $N'$ qubits with $poly(N')$ generic two-qubit gates, we can transform it into a circuit $U$ on a chain of $N=poly(N')$ qubits with nearest neighbor gates $W$ (with control on the left) and $S$ without loss of universality. We then add identity to our gate set and further transform the circuit $U$ to have the following form (see Figure \ref{CircuitFigure}). Rewrite the circuit as $K$ rounds of nearest neighbor gates $U_{k,g} \in \{W,S,I\}$,
where gate $U_{k,g}$ belongs to the $k$-th round and acts on the pair of qubits $w_g,w_{g+1}$:
\begin{eqnarray}
	U = (U_{K,N-1}\dots U_{K,1}) \cdots (U_{1,N-1}\dots U_{1,1}).
	\label{circuitsequence}
\end{eqnarray}

\begin{figure}
	\begin{center}
	\includegraphics[width=3.6in]{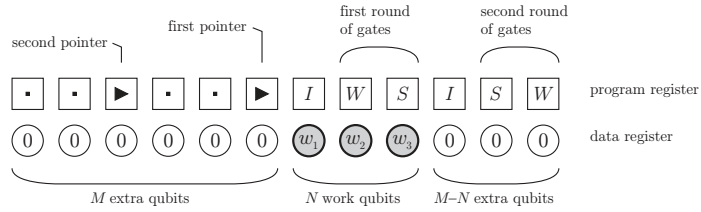}
	\caption{The initial state $\ket{\varphi}$ of the qudit chain for the circuit in Figure \ref{CircuitFigure}a).}
	\label{InitFigure}
	\end{center}
\end{figure}

We wish to encode the progression of the circuit $U$ into the states of a chain of qudits with dimension $10$, with length $L=poly(N)$.
The basis states of each qudit $\ket{q} = \ket{p}\otimes \ket{d}$ are 
constructed as a tensor product of a $5$-dimensional program register 
and a $2$-dimensional data register, where $p \in\{ \bul,\gat,W,S,I\}$ and $d \in \{0,1\}$ label the corresponding basis states.
We start by writing the initial product state 
\begin{eqnarray}
	\kets{\varphi} = \bigotimes_{j=1}^{L} \left(\ket{p_j} \otimes \ket{d_j}\right)_j,
	\label{start10product}
\end{eqnarray}
with $p_j$ and $d_j$ as in Figure \ref{InitFigure}
(here we give an example for the circuit in Figure \ref{CircuitFigure}a):
\begin{eqnarray}
	\begin{array}{c|cccccccccccc}
		j  &   1   &   & & & \cdots
		               &  M    & \cdots	
		              & & & &  & 2M \\
		\hline
		p_{j}  & \bul & \bul & \gat & \bul & \bul & \gat &
		\iga & \wga & \sga & 
		\iga & \sga & \wga 
		\\		
		d_{j}  & 0      & 0    & 0    & 0    & 0    & 0    & 
		w_1  & w_2  & w_3  &
		0    & 0    & 0        		
 	\end{array} \label{startstate10}
\end{eqnarray}
The qudit chain has length 
\begin{eqnarray}
	L = 2M = 2KN,
\end{eqnarray}
where $K$ is the number of the rounds of gates in \eqref{circuitsequence}.
The left half of the top (program) register contains $K$ pointer symbols $\gat$ at positions $kN$ for $k=1\dots K$ and empty symbols $\bul$ everywhere else. The right half holds the program in the form
\begin{eqnarray}
	I \underbrace{U_{1,1} \dots U_{1,N-1}}_{\textrm{first round of gates}}
	I \underbrace{U_{2,1} \dots U_{2,N-1}}_{\textrm{second round of gates}}
	I \dots
	I \underbrace{U_{K,1} \dots U_{K,N-1}}_{\textrm{last round of gates}}\, ,
	\label{program10}
\end{eqnarray} 
with $U_{k,g} \in \{W,S,I\}$ and each round of gates is preceded by an identity gate. Later we will show a way to execute the program while moving it to the left above the data. The bottom (data) register contains $N$ work qubits labeled $w_n$ in a canonical basis product state at positions $M+n$ for $n=1\dots N$. We designate $w_N$ as the readout qubit. Finally, the rest of the data register (positions $1,\dots,M$ and $M+N+1,\dots,2M$) is filled with extra qubits in the state $\ket{0}$. 

We now describe how to generate the set of states $\{\ket{\varphi_\sigma}\}$ encoding the progression of the quantum computation $U$. Starting from the initial state $\ket{\varphi}$ \eqref{startstate10}, we obtain each state $\ket{\varphi_\sigma}$ by applying a particular sequence $\sigma$ of the following two update rules. Each of these rules can be applied to two neighboring qudits in the chain. The first rule is: if there is an empty spot $\bul$ to the left of a symbol $A\in\{W,S,I\}$ (from now on we call it a `gate') somewhere in the program register, move $A$ one step to the left, leaving the data register unchanged.
\begin{eqnarray}
	\begin{array}{rccc}
	1\,:& 
		\four{\bul}{\aga}{x}{y}
		&\goes&
		\four{\aga}{\bul}{x}{y}
	\end{array}	\label{rule10x1}
\end{eqnarray}
The second rule says: at a place in the qudit chain where a gate $A\in \{W,S,I\}$ meets a `pointer' $\gat$, switch their positions and apply the gate $A$ to the qubits in the data register below.
\begin{eqnarray}
	\begin{array}{rccc}
	2\,:& 
		\four{\gat}{\aga}{x}{y}
		&\goes&
		\flour{\aga}{\gat}{A(x,y)}
	\end{array}	\label{rule10x2}
\end{eqnarray}
There is only one place where one of these rules can be applied to the initial state \eqref{startstate10} -- the second rule can be applied at the place where $\gat$ stands to the left of $I$. After this first step, there are several ways to proceed, generating many different states $\ket{\varphi_\sigma}$. 

\begin{figure}
	\begin{center}
	\includegraphics[width=3.2in]{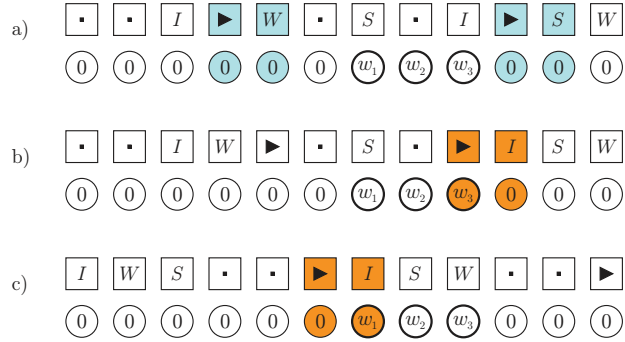}
	\caption{Analysis of the second rule \eqref{rule10x2}. a) A gate meeting a pointer above two extra qubits in the state $\ket{00}$. b) The gate meeting a pointer above the right boundary of the work qubits is the identity gate. c) The only gate meeting a pointer above the left boundary of the work qubits is the identity gate.}
	\label{ProblemFigure}
	\end{center}
\end{figure}

Let us look more closely at what happens to the initial state $\ket{\varphi}$ as we apply some sequence of rules. There are $K$ pointers $\gat$ in $\ket{\varphi}$, one for each round of gates in the circuit (see Figure \ref{InitFigure}). Observe that  $\ket{\varphi}$ is constructed in such a way that as a gate $U_{k,g}$ from the $k$-th round of gates moves to the left, it meets the $k$-th (counting from the right) pointer $\gat$ exactly above the work qubits $w_g,w_{g+1}$ (see Figure \ref{ProblemFigure}c for a depiction of the second pointer meeting the second round of gates). As noted before \eqref{circuitsequence}, this is the pair of qubits the gate $U_{k,g}$ should act on. It could then seem that after many applications of the update rules  the program moves to the left of the work qubits while the corresponding states $\ket{\varphi_\sigma}$ we obtain have the computation executed on their work qubits as planned.
However, before drawing that conclusion, we need to consider what happens when a gate $U_{k,g}$ from the $k$-th round meets a pointer while not above the work qubits (see Figure \ref{ProblemFigure}). First, if this happens above a pair of (extra) data qubits in the state $\ket{00}$, the qubits stay unchanged after we apply \eqref{rule10x2}, as we have $U_{k,g} \ket{00} = \ket{00}$ for the three possible $U_{k,g}$ (the controlled gate $W$, a swap or the identity). The second possibility is that a gate meets a pointer above the boundary of the work qubits, i.e. either above $d_M,w_1$ or above $w_N,d_{M+N+1}$.
Observe that the identity gates we inserted in front of each round of gates in \eqref{program10} are the only ones for which this can happen. The extra qubit and the work qubit involved then stay unchanged after the application of \eqref{rule10x2}. The extra qubits in the data register thus always remain in the state $\ket{0}$ and the only processing in the data register happens on the work qubits. The gates in \eqref{program10} are applied sequentially from left to right. Therefore, when all the gates move to the left half of the chain, the corresponding state $\ket{\varphi_{\sigma}}$ then contains the result of the quantum circuit $U$ in the state of the work qubits.

\subsubsection{The Hamiltonian}

We now construct a Hamiltonian whose transition rules will be (\ref{rule10x1}, \ref{rule10x2}) and their inverses
\begin{eqnarray}
	\begin{array}{rccc}
	1^\dagger\,:& 
		\four{\aga}{\bul}{x}{y}
		&\goes&
		\four{\bul}{\aga}{x}{y}
	\end{array},
	\qquad
	\begin{array}{rccc}
	2^\dagger\,:& 
		\four{\aga}{\gat}{x}{y}
		&\goes&
		\flour{\,\gat\,}{\aga}{A^\dagger(x,y)}
	\end{array}	\label{rule10x2i}.
\end{eqnarray}
Note that the inverse of the rule \eqref{rule10x2} involves applying $A^\dagger$ to the data qubits, uncomputing the gate. 
Our Hamiltonian is a sum of translationally invariant terms
\begin{eqnarray}
	H_{10} = - \sum_{j=1}^{L-1} \left( R+R^\dagger
		\right)_{(j,j+1)},
		\label{ourH10}
\end{eqnarray}
where $R$ corresponds to the rules (\ref{rule10x1}, \ref{rule10x2}) and acts on two neighboring qudits as
\begin{eqnarray}
	R = 
		\sum_{A\in\{W,S,I\}}
		\left[
			 \ket{\aga \bul}\bra{\bul \aga}_{p_1, p_2}  
			 \otimes 
			 \ii_{d_1,d_2}
		+
			 \ket{\aga\gat}\bra{\gat\aga}_{p_1,p_2}		
			 \otimes
			 A_{d_1,d_2}
		 \right],
\end{eqnarray}
where $p$ stands for the program register and $d$ for the data register of the respective qudit. 
Similarly, $R^\dagger$ corresponds to the inverse rules given in \eqref{rule10x2i}.

\subsubsection{The computational subspace}
\label{relevantsection}

To show that we can obtain a result of a quantum computation by time evolving the state \eqref{startstate10} with this Hamiltonian, we return to the set of states $\{\ket{\varphi_\sigma}\}$ generated using the rules (\ref{rule10x1}, \ref{rule10x2}).
There are only finitely many possible sequences $\sigma$. However, when we expand the ruleset by adding the inverse rules \eqref{rule10x2i}, the number of possible sequences $\sigma$ of rule (forward and backward) applications then becomes infinite.
For example, we get the possibility of returning back to the initial state $\ket{\varphi}$, completely undoing the computation. 
Nevertheless, the space of states $\{\ket{\varphi_{\sigma}}\}$ remains finite-dimensional for the following three reasons. 

First, the rules (\ref{rule10x1}, \ref{rule10x2}, \ref{rule10x2i}) do not change the order of the gates in the program register. Second, once we know the locations of the gates,
the rest of the program register is uniquely determined. It contains the sequence 
\begin{eqnarray}
	\underbrace{\bul \bul \cdots \bul}_{N-1} \gat
	\label{progfill}
\end{eqnarray}
repeatedly written out $K$ times from left to right at the positions unoccupied by the gates. Third, we constructed the initial state and the rules so that the gates in \eqref{program10} are applied (and uncomputed) to the work qubits sequentially,
while all the extra qubits remain in the state $\ket{0}$. 
Given the positions of the gates in the program register of $\ket{\varphi_\sigma}$, we can simply determine how many of the gates $U_{k,g}$ in \eqref{program10} have presently been applied to the work qubits. 
For any $\ket{\varphi_\sigma}$, there can be at most one pointer $\gat$ directly above the work qubits $w_1\dots w_N$.
\begin{enumerate}
\item If there is no pointer located above the work qubits (see Figure \ref{InitFigure}, \ref{ProblemFigure}a and \ref{ProblemFigure}c), label $k_p$ the number of pointers already to the right of $w_N$. The first $N k_p$ gates then have already been applied. 
\item If there is a pointer above the work qubits (see Figure \ref{ProblemFigure}b), all the gates to the left of this pointer have already been applied. 
\end{enumerate}
Let us look at a few examples of the first case. In Figure \ref{InitFigure}, $k_p=0$ and thus no gates have been applied yet. In Figures \ref{ProblemFigure}a) and \ref{ProblemFigure}c) $k_p=1$ and because $N=3$, we know that the first $3$ gates ($IWS$) have been executed. The state of the work qubits is thus equal to $S_{23} W_{12} \ket{w_1 w_2 w_3}$, in both Figure \ref{ProblemFigure}a) and \ref{ProblemFigure}c).

Therefore, regardless of the particular sequence of update rules (and their inverses) through which we obtained $\ket{\varphi_\sigma}$, the state of the data register is uniquely determined by the positions of the gates in the program register. In fact, all the states $\ket{\varphi_\sigma}$ with the same position of gates in the program register {\em are} the same state. Let us then label this state $\ket{\varphi_C}$, where $C$ is a weight $M$ bit string of length $L$. The $M$ ones in the string $C$ are located at positions
\begin{eqnarray}
	\{a_1^{(C)},\dots,a_M^{(C)}\},
\end{eqnarray}
corresponding to the positions of the $M$ gates in the program register of the state $\ket{\varphi_\sigma}$.
The number of different states in the set $\{\ket{\varphi_C}\}$ 
is ${L\choose M}$. Note also that the states $\ket{\varphi_C}$ with different $C$ are mutually orthogonal, as their corresponding program registers are in mutually orthogonal product states. The set $\{\ket{\varphi_C}\}$ thus forms a basis of an ${L\choose M}$-dimensional subspace of the Hilbert space of the qudit chain.

The transition rules in the Hamiltonian \eqref{ourH10} correspond to (\ref{rule10x1}, \ref{rule10x2}, \ref{rule10x2i}). The time evolution of the initial state $\ket{\varphi}$ \eqref{startstate10} will thus necessarily happen within
\begin{eqnarray}
	\cH_{comp} = span \{ \ket{\varphi_C} \},
	\label{Hcomp}
\end{eqnarray}
the subspace spanned by the states $\ket{\varphi_C}$. 
We call $\cH_{comp}$ the computational subspace.
In Section \ref{fermionsection}, we show that our Hamiltonian \eqref{ourH10} restricted to the computational subspace has a particularly simple form corresponding to a Hamiltonian of a fermionic system. This will allow us to analyze the time evolution of the initial state.


\subsection{Required Time Analysis}
\label{d10time}

Our model of computation with the Hamiltonian \eqref{ourH10} consists of initializing the qudit chain in the state $\kets{\varphi}$ \eqref{startstate10} and evolving the system for a time $\tau$ randomly chosen between $0$ and $\tau_{10}$. 
In this Section, we will show that when we measure the output qubit $w_N$ in the data register at time $\tau$, we will read out the result of the quantum computation $U$ with high probability. Moreover, we will show that the upper bound on the running time of our model, $\tau_{10}$, is polynomial in $L$.

Each state $\ket{\varphi_{C}}$ described above can be written as
\begin{eqnarray}
	\ket{\varphi_{C}} &=& 
	\kets{\pi_C}_{program} \otimes
		 \kets{\theta_C}_{work} \otimes \kets{\alpha}_{extra},
		 \label{stateCdecompose}
\end{eqnarray}
where $\kets{\pi_C}_{program}$ labels a state of the program register of the chain, the corresponding state of the work qubits is $\kets{\theta_C}_{work}$ and the extra data qubits are in the state $\ket{\alpha}_{extra} = \kets{0}_{extra}^{\otimes (2M-N)}$. Note that while each program register state $\kets{\pi_C}_{program}$ is a product state, the corresponding state of the work qubits $\kets{\theta_C}_{work}$ is entangled. 

The time evolved state $\kets{\varphi(\tau)}$ is a superposition of the states $\ket{\varphi_{C}}$:
\begin{eqnarray}
	\kets{\varphi(\tau)} = e^{-iH_{10} \tau} \ket{\varphi} 
	= 
		\sum_{C} c_{C}(\tau) 
		\,\kets{\varphi_C}
		= 
		\sum_{C} c_{C}(\tau) 
		\,\kets{\pi_C}_{program} \otimes
		 \kets{\theta_C}_{work} \otimes \kets{\alpha}_{extra},
		 \label{evolved10a}
\end{eqnarray}
where the sum over $C$ is over all weight $M$ length $L$ bit strings and $c_C(\tau)$ is the amplitude of the state $\ket{\varphi_C}$ in $\kets{\varphi(\tau)}$.

When all of the gates in $\ket{\pi_C}_{program}$ have moved to the left of the work qubits, the state of the work qubits holds the output state of the computation $U$. For our original setup \eqref{startstate10}, there is only one such state:
\begin{eqnarray}
	\ket{\varphi_{111111000000}} = \ket{\iga \wga \sga \iga \sga \wga \bul \bul \gat \bul \bul \gat}_{program} \otimes
		 U\ket{w_1 \dots w_N}_{work} \otimes
		 \ket{\alpha}_{extra}.
\end{eqnarray}
For our model of computation, this would mean that we would need a single amplitude $c_{1\dots10\dots0}(\tau)$ to be large at a random time $\tau$, which is unlikely. 
However, a simple modification will turn the tables.
Let us pad the qudit chain by $(f-1)M$ extra sites on the left and $M$ extra sites on the right, with $f$ a constant to be determined later. Let the new initial state be 
\begin{eqnarray}
	\kets{\varphi^{pad}} &=& 
		\begin{array}{|c|}
			\hline \bul\\
			\hline 0\\
			\hline
		\end{array}^{\,\,\otimes (f-1)M}
		\otimes \kets{\varphi} \otimes 
		\begin{array}{|c|}
			\hline\iga\\
			\hline 0\\
			\hline
		\end{array}^{\,\,\otimes M}\, ,
	\label{startstate10padded}
\end{eqnarray}
where $\ket{\varphi}$ is the original initial state \eqref{startstate10}.
The original chain had length $2M$, so the length of this padded chain is $L=(f+2)M$, while the number of gates becomes $2M$.
As in Section \ref{relevantsection}, let 
us now label $\{\ket{\varphi_D}\}$ the set of states which we can obtain from \eqref{startstate10padded} by applying rules (\ref{rule10x1}, \ref{rule10x2}, \ref{rule10x2i}). This time, the label $D$ is a weight $2M$ length $L=(f+2)M$ bit string, and $\{a_1^{(D)},\dots,a_{2M}^{(D)}\}$ denote the positions of the $2M$ gates in the program register of $\ket{\varphi_D}$.

The $2M$ gates in the program register of the initial state $\kets{\varphi^{pad}} = \kets{\varphi_{0\dots 01\dots 1}}$ are located at positions $a_m^{(0\dots01\dots1)} = fM+m$ with $m=1\dots2M$. Only the first $M$ gates are relevant for the computation and the other $M$ are the identity gates we inserted as padding.
These identity gates leave the data register intact, regardless of whether they move to the left of the work qubits or stay to the right of them. Therefore, every state $\ket{\varphi_{D}}$ in which the first $M$ (relevant) gates have already moved to the left of the work qubits contains the finished computation in the state of its work qubits. Let us label the corresponding set of weight $2M$ length $L=(f+2)M$ bit strings 
\begin{eqnarray}
	\mathcal{D}_{done} = \{D: a_{M}^{(D)}\leq fM\}.
\end{eqnarray}
For all states $\ket{\varphi_{D}}$ with $D\in \mathcal D_{done}$, the state of the work qubits is the same and equal to $\ket{\theta_U} = U \ket{w_1 \dots w_N}$.
Let us write the time evolved state of the modified chain using the notation  
\eqref{stateCdecompose} as  
\begin{eqnarray}
	\kets{\varphi^{pad}(\tau)} &=& e^{-iH_{10} \tau} \kets{\varphi^{pad}} \nonumber\\
		&=&	\left( 
			\sum_{D\notin \mathcal{D}_{done}} c_{D}(\tau) 
			\,\kets{\varphi_D} 
			\right)
		+
			\left( 
			\sum_{D\in \mathcal{D}_{done}} c_{D}(\tau) 
			\,\kets{\pi_D}_{program} 
			\right)
			\otimes
			 \kets{\theta_U}_{work} \otimes 
			 \kets{\alpha}_{extra}
		 \label{evolved10b} \\
		&=&	\sqrt{1-p_{10}} \ket{\varphi_{fail}(\tau)}
		+
			\sqrt{p_{10}} \ket{\pi_{done}(\tau)} \otimes
			 \kets{\theta_U}_{work} \otimes 
			 \kets{\alpha}_{extra}
			. \label{evolved10c}
\end{eqnarray}
The state of the work qubits with the computation done $\ket{\theta_U}_{work}$ factorizes out in the second term. The two terms are orthogonal, as the states of the program register of $\ket{\varphi_{D}}$ with different $D$ are mutually orthogonal. 
We will now analyze the amplitude of the second term. Assume we have a measurement discriminating different states $\ket{\pi_D}$ of the program register. In that case $p_{10}$ denotes the probability of obtaining $\ket{\pi_D}$ with $D\in \mathcal{D}_{done}$
when measuring at a random time $\tau \leq \tau_{10}$. 

\subsubsection{Mapping onto a fermionic system}
\label{fermionsection}

We will now prove that when we choose the time $\tau$ uniformly at random in $(0,\tau_{10})$, with $\tau_{10} = poly(M)$, the probability of finding a state with the computation executed (with $a_M\leq fM$) is $p_{10} \geq \frac{5}{6} -  O\left(\frac{L}{\tau_{10}}\right)$
with $L=(f+2)M$.
For this, we will analyze the time evolution of $\kets{\varphi^{pad}}$ \eqref{startstate10padded} with $H_{10}$ \eqref{ourH10}. 
We will do it by mapping the states of the qudit chain onto states of a chain of spin-$\half$ particles, where the positions of gates in the qudit chain will correspond to fermionic excitations in the spin chain. By analyzing the latter system, we will bound $p_{10}$.

As in \ref{relevantsection}, the time evolution of the padded initial state \eqref{startstate10padded} happens only in the computational subspace 
\begin{eqnarray}
	\mathcal{H}_{comp}^{pad} = span \{\ket{\varphi_D}\}.
	\label{Hcomppad}
\end{eqnarray} 
The basis states of the computational subspace have the form
\begin{eqnarray}
	\ket{\varphi_D} = \ket{\pi_{D}}_{program} 
		\otimes 
			\left( U^{(D)} \ket{w_1 \dots w_N}_{work} \right) 
			\otimes 			
			\ket{\alpha}_{extra},
	\label{compspan}
\end{eqnarray}
where the label $D$ is a weight $2M$ bit string of length $L=(f+2)M$. The states $\ket{\pi_{D}}$ of the program register are canonical basis product states with 
\begin{eqnarray}
	I U_{1,1} \dots U_{1,N-1}\,
	I U_{2,1} \dots U_{2,N-1}\,
	\cdots\,
	I U_{K,1} \dots U_{K,N-1}
	\,\underbrace{II \dots I}_{M}
	\label{program10pad}
\end{eqnarray}
written out from left to right at the positions $\{a_1^{(D)},\dots,a_{2M}^{(D)}\}$ of the ones in $D$, and the rest of the program register is filled with 
\begin{eqnarray}
	\underbrace{\bul \bul \cdots \bul}_{N-1} \gat
	\label{fill10pad}
\end{eqnarray}
written out $fK$ times from left to right. Finally, $U^{(D)}$ stands for the part of the sequence of unitaries \eqref{program10pad} that has already been applied to the work qubits when the gates are in positions $\{a_1^{(D)},\dots,a_{2M}^{(D)}\}$. We described the procedure to determine $U^{(D)}$ from $D$ in Section \ref{relevantsection}.

We will now focus on the restriction of $H_{10}$ to $\mathcal{H}_{comp}^{pad}$, as the time-evolved state \eqref{startstate10padded} never leaves the computational subspace.
This restriction has a simple form---it is the negative of an adjacency matrix, where two states are connected if one can be obtained from the other by one of the rules (\ref{rule10x1}, \ref{rule10x2}, \ref{rule10x2i}). More explicitly, two of the states $\ket{\varphi_D}$ are connected, if one can obtain the other by having one gate hop to the left (or right) and applying the appropriate unitary (or do nothing) to the work qubits. 
Consider now
\begin{eqnarray}
	H_q = - \sum_{j=1}^{L-1} (\ket{10}\bra{01}+\ket{01}\bra{10})_{j,j+1},
	\label{spinham}
\end{eqnarray}
the Hamiltonian of a line of $L=(f+2)M$ spin-$\half$ particles with a simple hopping interaction, restricted to the subspace spanned by states with $2M$ spin up particles.
Observe that 
with the mapping 
\begin{eqnarray}
	\ket{\varphi_D}_{\textrm{qudit chain}} \quad \leftrightarrow \quad \ket{D}_{\textrm{spin chain}},
\end{eqnarray}
where $D$ is a weight $2M$ bit string of length $L$,
the restriction of $H_{10}$ \eqref{ourH10} to $\cH_{comp}^{pad}$ \eqref{Hcomppad} is matrix-wise the same as the restriction of $H_q$ to the subspace with $2M$ up spins, i.e. 
\begin{eqnarray}
	H_{10} |_{\cH_{comp}^{pad}} &=& H_q |_{2M\,\textrm{up spins}}.
\end{eqnarray}
To show that $p_{10}$ in \eqref{evolved10c} is large, it will be convenient to analyze the time evolution of the initial state of the spin chain
\begin{eqnarray}
	\kets{\underbrace{0\dots0}_{fM}\underbrace{1\dots1}_{2M}}
\end{eqnarray}
with the spin Hamiltonian \eqref{spinham}
instead of the time evolution of \eqref{startstate10padded} with \eqref{ourH10}.
First, we use the usual Wigner-Jordan transformation to define the operators
\begin{eqnarray}
	b^{\dagger}_j &=& \sigma^{z}_1 \dots \sigma^{z}_{j-1} \otimes 
		\ket{1}\bra{0}_j \otimes \ii_{j+1, \dots, L}, 
			\label{wigner1}\\ 
	b_j &=& \sigma^{z}_1 \dots \sigma^{z}_{j-1} \otimes 
		\ket{0}\bra{1}_j \otimes \ii_{j+1, \dots, L}. 
			\label{wigner2}
\end{eqnarray}
We invite the reader to verify the properties $\{b_i,b^{\dagger}_j\} = \delta_{ij} \ii$ and $b_j^2 = b_j^{\dagger 2} = 0$ for $b^{\dagger}_j$ and $b_j$. Thus, $b^{\dagger}_j$ and $b_j$ can be viewed as the creation and annihilation operators for a fermion at site $j$. Rewriting \eqref{spinham} in terms of \eqref{wigner1} and \eqref{wigner2}, we obtain 
\begin{eqnarray}
	H_f = - \sum_{j=1}^{L-1} b^{\dagger}_j b_{j+1} + h.c.,
	\label{Hf}
\end{eqnarray}
a Hamiltonian for a system of free fermions in second quantization.
Following our mapping, the initial state $\kets{\varphi^{pad}}$ of the qudit chain thus corresponds to the state of the fermionic system $\ket{\Psi} = b^{\dagger}_{fM+1} \dots b^{\dagger}_{(f+2)M} \ket{0}$ with $2M$ fermions on the right end of the line (here $\ket{0}$ is the state with no fermions).
We now use the following Lemma (proved in Appendix \ref{d10proof}):

\begin{lem3}
	Consider the state $\ket{\Psi}$ of $2M$ fermions on the right end of a line with $L=(f+2)M$ sites. Let the system evolve for a time chosen uniformly at random between $0$ and $\tau_{10}$ with the Hamiltonian given by \eqref{Hf} and measure the number of fermions in the region $1\leq x \leq fM$. The probability to measure a number greater than $M$
is $p_{10} \geq \frac{f-2}{f+2} - O\left(\frac{L}{\tau_{10}}\right)$.
\end{lem3}

Let us choose $f=22$ and $\tau_{10} = O(L\log L) = O(M\log M)$. Following the mapping we did from our qudit chain backwards, this implies that when we initialize the qudit chain of length $L=24M$ in $\kets{\varphi^{pad}}$ as in \eqref{startstate10padded} and let it evolve with $H_{10}$ \eqref{ourH10} for a random time 
$\tau\leq \tau_{10}$,
the probability for the chain to be in a state where the gate particles have moved sufficiently to the left for the computation to be done ($a_M \leq fM$) is 
\begin{eqnarray}
	p_{10} \geq \frac{5}{6} - O\left(\frac{1}{\log M}\right).
\end{eqnarray}  
Using this bound on the probability of having the computation done
we now proceed to prove universality of our model for the class BQP.

\subsubsection{Universality for BQP}

To prove the universality of our model for recognizing languages in the class BQP, let us recall its definition. Consider a language $L$ in BQP, a uniform family of circuits $U$ and a problem instance $x$. When $x\in L_{yes}$,
the set of instances with the answer `yes', the probability of the circuit $U$ outputting `yes' is not smaller than $\frac{2}{3}$. On the other hand, when $x\notin L_{yes}$, the probability of the circuit outputting `yes' is not greater than $\frac{1}{3}$.
Let us assume the worst case for our circuit $U$, i.e. that the circuit outputs `yes' on a good proof for a `yes' instance with probability $p_U = \frac{2}{3}$.
In the language of spins, the circuit $U$ outputs `yes' when we measure spin up on the output qubit. Therefore, the expected value of measuring $\sigma_{z}$ on output qubit of the circuit $U$ is bounded from below by
\begin{eqnarray}
	\langle \sigma^{(z)}_{w_N} \rangle^{circuit}_{yes} \geq 
	1\times p_U + (-1)\times(1-p_U) = 2p_U-1 = \frac{1}{3}
	\label{yesoutput}
\end{eqnarray}
when $x\in L_{yes}$. Analogously, for $x\in L_{no}$, it is bounded from above by
\begin{eqnarray}
	\langle \sigma^{(z)}_{w_N} \rangle^{circuit}_{no} \leq 
	-2p_U + 1
	= - \frac{1}{3}.
	\label{nooutput}
\end{eqnarray}

To solve BQP problems with our automaton, we need to distinguish the `yes' from the `no' cases, i.e. we need to show that the expectation value of measuring $\sigma^{(z)}$ on the output qubit of our automaton at a random time $\tau\leq \tau_{10}$ is greater than zero in the `yes' case, and smaller than zero in the `no' case. 
The probability of finding a state where the computation is finished is $p_{10}$. Let us consider a `yes' instance ($x\in L_{yes}$).
Using \eqref{evolved10c} and recalling that the two terms in it are orthogonal, we have
\begin{eqnarray}
	\langle \sigma^{(z)}_{w_N} \rangle_{yes} =
	\bra{\varphi^{pad}(\tau)} \sigma^{(z)}_{w_N} 	\ket{\varphi^{pad}(\tau)}
		&=& 
			(1-p_{10}) 
				\bra{\varphi_{fail}(\tau)}\sigma^{(z)}_{w_N} 
				\ket{\varphi_{fail}(\tau)}
			+	p_{10} 
			\underbrace{
				\bra{\theta_U} 
				\sigma^{(z)}_{w_N} 
				\ket{\theta_U}_{work} 
			}_{\textrm{output of }U}.
		\label{expect10}
\end{eqnarray}
The second term is the circuit output \eqref{yesoutput}, therefore
\begin{eqnarray}
	\bra{\theta_U} 
			\sigma^{(z)}_{w_N} 
			\ket{\theta_U}_{work}
	=
	\langle \sigma^{(z)}_{w_N} \rangle^{circuit}_{yes}
	\geq 
		2p_U - 1.
\end{eqnarray}
The first term can be bounded from below (adversarially, i.e. for every time the computation is not finished, the output qubit gives the opposite of the correct answer) by
\begin{eqnarray}
	\bra{\varphi'}\sigma^{(z)}_{w_N}	\ket{\varphi'} \geq -1. 
\end{eqnarray}
Putting it together, we have
\begin{eqnarray}
	\langle \sigma^{(z)}_{w_N} \rangle_{yes} 
		&\geq& p_{10} (2p_U - 1) - (1-p_{10})
		= 2p_{10}p_{U} - 1.
	\label{expect10done}
\end{eqnarray}
Analogously, for the $x\notin L$ case, we obtain
\begin{eqnarray}
	\langle \sigma^{(z)}_{w_N} \rangle_{no} 
		&\geq& - 2p_{10}p_{U} + 1.
	\label{expect10doneNO}
\end{eqnarray}

Therefore, equations \eqref{expect10done} and \eqref{expect10doneNO} now read
\begin{eqnarray}
	\langle \sigma^{(z)}_{w_N} \rangle_{yes} 
		&\geq& \phantom{-}\frac{1}{9} - O\left(\frac{1}{\log M}\right), \nonumber
		\\
	\langle \sigma^{(z)}_{w_N} \rangle_{no} 
		&\leq& -\frac{1}{9} + O\left(\frac{1}{\log M}\right).
	\label{sigmayesno}
\end{eqnarray}
Therefore, we can recognize any language in BQP using the HQCA we described above.

As an aside, note that there is a way to determine that we obtained a state in which the computation has been done with certainty (and thus getting rid of the first term in \eqref{expect10}). We could have chosen to measure all the program qudits to the right of the first work qubit and check whether all the $S$ and $W$ are gone. This happens with the above probability $p_{10}$, and the postselected state of the work qubits now surely contains the output of the circuit $U$. Note also that we can think of the state of all the work qubits as the circuit output, as compared to only the last work qubit. Nevertheless, thinking only about the last work qubit is enough to ensure universality of our HQCA for the class BQP.


Concurrently with our preprint, Chase and Landahl \cite{ChaseLandahl} found a $d=8$ Hamiltonian computer construction universal for BQP. However, their Hamiltonian is not translationally invariant, and thus not a HQCA. If we release the translational invariance in our $d=10$ HQCA model, we can be sure where the computational qubits are and use the gate set $\{S,SW\}$ instead of $\{S,W,I\}$ in our program register. This brings the required dimensionality of our model to $d=8$ as well. However, it remains an open question whether this can be done with keeping the translational invariance of the Hamiltonian, i.e. whether a $d\leq 10$ HQCA in 1D exists.


\section{The $d=20$ HQCA}
\label{d20section}

We now present our second construction, a HQCA for a chain of $20$-dimensional qudits.
As in Section \ref{d10construct}, we describe an encoding of the progression of a quantum circuit $U$ into a set of states of a qudit chain. However, the geometry of this set of states $\ket{\psi_t}$ will be now much simpler, as we can label them by a ``time'' label $t=1,\dots, T$ with $T=poly(N)$, thinking of the set of states as a
``line''.
The Hamiltonian $H_{20}$ we construct induces a quantum walk on this ``line'' of states. We conclude by proving that when we let the initial state $\ket{\psi_0}$ evolve with $H_{20}$ for a time $\tau$
chosen uniformly at random between $0$ and $\tau_{20} = O(T\log T)$, we can read out the result of the quantum computation $U$ with probability $p_{20}\geq \frac{5}{6}-O\left(\frac{T}{\tau_{20}}\right)$ by measuring two of the qudits in the computational basis.

\subsection{The Construction}
\label{d20construct}

We encode the progression of a quantum circuit $U$ in the form \eqref{circuitsequence} (see also Figure \ref{CircuitFigure}) into a set of states $\ket{\psi_t}$ of a qudit chain with length $L = (2K-1)(N+1) + 2$. As in Section \ref{d10construct}, each qudit consists of a program register and a data register. 
The data register is again two-dimensional, but the program register can now be in the following $10$ states:
\begin{eqnarray}
	\wga,\sga,\iga &:& \textrm{the program sequence,} 
		\nonumber\\
	\wci,\sci,\ici &:& \textrm{marked characters in the program sequence, used to propagate} \nonumber\\
			&& \textrm{the active spot to the front (left) of the program sequence,} \nonumber\\
	\gat &:& \textrm{apply gate symbol,} \nonumber\\
	\mov &:& \textrm{shift program forward,} \nonumber \\
	\tur &:& \textrm{a turn-around symbol,} \nonumber \\
	\bul &:& \textrm{empty spot (before/after the program)}. \nonumber
\end{eqnarray}

Similarly to \eqref{start10product} and \eqref{startstate10},
the initial product state $\ket{\psi_0} = \bigotimes_{j=1}^{L} \left(\ket{p_j} \otimes \ket{d_j}\right)_j$
is given by 
(we write an example for the circuit in Figure \ref{CircuitFigure}a)
\begin{eqnarray}
	\begin{array}{c|cccccccccccccc}
		     j & 1 & \cdots     &      &      &      &      & 
		&&&&&& \cdots & L\\
		\hline
		p_j & \bul & \bul & \bul & \bul & \bul & \bul &
		\iga & \wga & \sga & 
		\iga & \iga & \sga & \wga & \tur 
		\\		
		d_j & 0    & 1    & 0    & 0    & 0    & 1    & 
		w_1  & w_2  & w_3  &
		1    & 0    & 0    & 0    & 1		
 	\end{array} \label{startstate}
\end{eqnarray}
In general, the data register contains $N$ work qubits (labeled $w_n$ in our example) at positions $(K-1)(N+1)+2+n$ for $n=1:N$ (counting from the left). Qubit $w_N$ is the designed output qubit for the computation, i.e. once the computation is done, $w_N$ contains the output of $U$. Next, the data register contains qubits in the state $\ket{1}$ at positions $(k-1)(N+1) + 2$ for $k=1\dots 2K$ and qubits in the state $\ket{0}$ everywhere else. The 1's serve as sequence boundary markers.
The program register has 
empty symbols $\bul$ on the left, and then it contains the program in the form 
\begin{eqnarray}
	I\, 
	\underbrace{U_{1,1} \dots U_{1,N-1}}_{1^{\textrm{st}}\textrm{ gate sequence}} 
	I\, I\, 
	\underbrace{U_{2,1} \dots U_{2,N-1}}_{2^{\textrm{nd}}\textrm{ gate sequence}} 
	I\, I\, 
	\:\cdots\:
	I\, I\, 
	\underbrace{U_{K,1} \dots U_{K,N-1}}_{\textrm{last gate sequence}},
	\label{program}
\end{eqnarray}
with the program written from left to right.
In our example \eqref{startstate}, the first gate sequence (see Figure \ref{CircuitFigure}a) is $WS$ and the second gate sequence is $SW$. Finally, the last qudit in the program register is in the state $\tur$, marking an active spot in the computation. 

We now give the rules to obtain the sequence of states $\ket{\psi_t}$ from $\ket{\psi_0}$. These rules are constructed so that there is always only one of them that can be applied to a given state $\ket{\psi_t}$, thus giving us a unique state $\ket{\psi_{t+1}}$.  
(Also, using the rules backwards, one obtains a unique $\ket{\psi_{t-1}}$ from $\ket{\psi_t}$). The first three are
\begin{eqnarray}
	\begin{array}{rccc}
	\raiseonebox 
	1\,:& 
		\band{\aga}{\tur} 
		&\goes&
		\band{\aci}{\bul} 
		\\
	\raiseonebox 
	2\,:& 
		\band{\aga}{\bci}
		&\goes&
		\band{\aci}{\bga}
		\\
	3\,:& 
		\band{\bul}{\aci}
		&\goes&
		\band{\tur}{\aga} 
	\end{array}
	\label{rule123}
\end{eqnarray}
where $A,B$ stands for either $W,S$ or $I$. 
These rules ensure the passing of the active spot from the back end (right side) of the program to the front (left side), without modifying the data register or the order of the gates in the program sequence. Next, we have
\begin{eqnarray}
	\begin{array}{rccccrccc}
	4a\,:&
		\triUR{\bul}{\tur}{1} 
		&\goes&
		\triUR{\bul}{\gat}{1}
	&\quad&
	4b\,:& 
		\triUR{\bul}{\tur}{0}
		&\goes&
		\triUR{\bul}{\mov}{0}
	\end{array} \label{rule4}
\end{eqnarray}
After the active spot has moved to the front of the program, there are two possibilities. The turn symbol $\tur$ can change to the apply gate symbol $\gat$ (rule 4a), or to the shift program symbol $\mov$ (rule 4b), depending on whether the data qubit below contains the sequence boundary marker state $1$. Afterward, for the states containing the apply gate symbol $\gat$, we have:
\begin{eqnarray}
	\begin{array}{rccccrccc}
	5a\,:& 
		\four{\gat}{\aga}{x}{y}
		&\goes&
		\flour{\aga}{\gat}{A(x,y)}
	&\quad&
	6a\,:& 
		\triUL{\gat}{\bul}{1}
		&\goes&
		\triUL{\tur}{\bul}{1}
	\end{array}	\label{rule5}
\end{eqnarray}
(Note that rule 6a at the right end of the chain involves only the two particles directly above each other, as no particle to the right of them exists.)
When applying rule 5a, the apply gate symbol $\gat$ moves 
to the right, while a gate from the program sequence is applied to the qubits in the data register below. 
Applying the rule repeatedly, the $\gat$ symbol moves to the right end of the program sequence. 
As an example, we now write out the state $\ket{\psi_{12}}$ that we obtained from the state $\ket{\psi_0}$ applying rules 1, 2 (6 times), 3, 4a and 5a (3 times) from the state $\ket{\psi_0}$. 
\begin{eqnarray}
	\ket{\psi_{12}} &=& \quad \Big[ \begin{array}{cccccccccccccc}
		\bul & \bul & \bul & \bul & \bul & \iga & 
		\wga & \sga & \gat & 
		\iga & \iga & \sga & \wga & \bul 
		\\		
		0    & 1    & 0    & 0    & 0    & 1    & 
		\multicolumn{3}{c}{\ket{\,\,\dots\theta\dots\,\,}}
		 &
		1    & 0    & 0    & 0    & 1		
 	\end{array} \Big], \label{psi12}
\end{eqnarray}
where $\ket{\dots\theta\dots}$ stands for the state of the three work qubits after the gates $W_{12}$ and then $S_{23}$ were applied to them.
Let us have a closer look at the marker qubits (all qubits in the data register except for the work qubits $w_n$) and the application of rule 5a. The marker qubits stay unchanged for all $\ket{\psi_t}$.
The gate applied to pairs $\ket{0}\ket{1}$ and $\ket{1}\ket{0}$ of marker qubits or the pairs of qubits $\ket{1}\ket{q_1}$ and $\ket{q_N}\ket{1}$ (the left and right ends of the work qubit sequence) is always $I$, because of the identity gates we inserted between sequences of gates in the program \eqref{program}.
Finally, the qubit pairs $\ket{0}\ket{0}$ between the 1 markers do not change under the swap operation or the $W$ gate (a controlled gate). 

After the apply gate $\gat$ symbol gets to the end of the sequence, it changes into the turn symbol $\tur$ via rule 6a. Note that the boundary markers in the data register are spaced in such a way, that the $\gat$ symbol will arrive at the right end of the sequence when the qubit below is in the state 1. Using rule 6a, $\gat$ will then change into the turn symbol $\tur$. 
 After applying rules 1, 2 (6 times) and 3, the active spot again moves to the left of the program. Because the $\tur$ symbol is now above a 0 marker qubit, rule 4b can be used, and we get a state with the shift program symbol $\mov$. Finally, here are the last two rules:
\begin{eqnarray}
	\begin{array}{rccccrccc}
	5b\,:& 
		\band{\mov}{\aga}
		&\goes&
		\band{\aga}{\mov}
	&\quad&
	6b\,:& 
		\triUL{\mov}{\bul}{0}
		&\goes&
		\triUL{\tur}{\bul}{0}
	\end{array}	\label{rule6}
\end{eqnarray}
where again $A$ stands for either $\wga,\sga$ or $\iga$. 
Rule 5b makes the program shift to the left while the $\mov$ symbol moves to the right. Finally, rule 6b deals with what happens when the $\mov$ symbol arrives at the end of the program sequence. Because of the way we constructed the data register in $\ket{\psi_0}$, the data
qubit below the $\mov$ symbol will then be in the state $0$, so that the $\mov$ symbol changes to the turn symbol $\tur$. The reason why we need to look at the qubit in the data register below the $\tur$ symbol in rules 6a and 6b is that when we apply the rules backwards (making $\ket{\psi_{t-1}}$ from $\ket{\psi_t}$), again only one of them applies for each $\ket{\psi_t}$.

After applying rule 1, 2a (6 times), 3 and 4b, the $\mov$ symbol appears again and starts shifting the program further to the left. After several rounds of this, when the program shifts to the left by $N+1$, rule 4a can be used again (as the $\tur$ symbol will be above a 1 marker qubit), and subsequently, the $\gat$ symbol facilitates the application of the second sequence of gates to the work qubits.

After many applications of the above rules, we arrive at the state $\ket{\psi_T}$, for which none of our (forward) rules apply. 
\begin{eqnarray}
	\ket{\psi_T} &=& \quad \Big[ \begin{array}{cccccccccccccc}
		\tur & \iga & \wga & \sga & \iga & \iga & 
		\sga & \wga & \bul & 
		\bul & \bul & \bul & \bul & \bul
		\\		
		0    & 1    & 0    & 0    & 0    & 1    & 
		\multicolumn{3}{c}{\ket{\,\,\dots\theta'\dots\,\,}}  &
		1    & 0    & 0    & 0    & 1		
 	\end{array} \Big]. \label{endstate}
\end{eqnarray}
This is the state in which the program has moved to the left of the qudit chain, and all sequences of gates have been applied to the qubits in the data register.
The state $\ket{\dots\theta'\dots}$ is thus the output state of the circuit $U$ and the last of the work qubits ($w_N$) holds the output of the quantum computation.

Starting from \eqref{startstate}, we have constructed the set of states $\ket{\psi_t}$ for $t=0\dots T$ with $T 
= O(K^2 N^2) = poly(N)$. As $t$ grows, these states encode the progress of a quantum circuit $U$. 
Let us now think of the geometry of this set of states. They are labeled by a discrete label $t$, with the state $\ket{\psi_{t}}$ 
obtainable only from the states $\ket{\psi_{t-1}}$ and $\ket{\psi_{t+1}}$ using the above rules and their backward applications.
Therefore, the states $\ket{\psi_t}$ can be thought of as position basis states on a line of length $T+1$
\begin{eqnarray}
	\ket{\psi_{t}} \quad \leftrightarrow \quad \ket{t}_{line},
	\label{d20map}
\end{eqnarray}
where $t=0\dots T$.

Let us choose a Hamiltonian $H_{20}$ for this system 
as a sum of translationally invariant terms:
\begin{eqnarray}
	H_{20} = - \sum_{i=1}^{L-1} 
		\sum_{k=1}^{6b} \left( P_{k} + P_{k}^{\dagger} \right)_{(i,i+1)}
		\label{ourH}
\end{eqnarray}
where the terms $P_k$ correspond to the rules 1-6b \eqref{rule123},\eqref{rule4},\eqref{rule5} and \eqref{rule6} and act on two neighboring qudits as 
\begin{eqnarray}
	P_1 &=& 
		\sum_{A\in\{W,S,I\}}			
			\ket{\aci\bul}\bra{\aga\tur}_{p_1,p_2}
		 \otimes 
		\ii_{d_1,d_2}	
			, \\
	P_2 &=& 
		 \sum_{A,B\in\{W,S,I\}} 
		\ket{\aci\bga}\bra{\aga\bci}_{p_1,p_2}
		\otimes
		\ii_{d_1,d_2} 
		,	 \\
	P_3 &=& 
		 \sum_{A\in\{W,S,I\}}
			\ket{\tur\aga}\bra{\bul\aci}_{p_1,p_2}
		\otimes	\ii_{d_1,d_2}, 
\end{eqnarray}
and
\begin{eqnarray}
	P_{4a} &=& 
			\ket{\bul\gat}\bra{\bul\tur}_{p_1,p_2}
		  \otimes 
			\ii_{d_1}
		  \otimes 
			\ket{1}\bra{1}_{d_2}, \\
	P_{4b} &=& 
		\ket{\bul\mov}\bra{\bul\tur}_{p_1,p_2}
			\otimes
			\ii_{d_1}\otimes \ket{0}\bra{0}_{d_2}, \\
	P_{5a} &=& 
		\sum_{A\in\{W,S,I\}}  
			\ket{\aga\gat}\bra{\gat\aga}_{p_1,p_2}
			\otimes
			A_{d_1,d_2}, \\
	P_{5b} &=& 
		\sum_{A\in\{W,S,I\}} \ket{\aga\mov}\bra{\mov\aga}_{p_1,p_2}
			\otimes
			\ii_{d_1,d_2}, \\
	P_{6a} &=& 
		\ket{\tur\bul}\bra{\gat\bul}_{p_1,p_2}
			\otimes
			\ket{1}\bra{1}_{d_1} \otimes \ii_{d_2}, \\
	P_{6b} &=& 
			\ket{\tur\bul}\bra{\mov\bul}_{p_1,p_2}
			\otimes
			\ket{0}\bra{0}_{d_1} \otimes \ii_{d_2}.
\end{eqnarray}
When thinking of the set of states $\ket{\psi_t}$ as the set of positions of a particle on a line \eqref{d20map}, $H_{20}$  becomes 
\begin{eqnarray}
	H_{line} = - \sum_{t=0}^{T-1} \big(
		\ket{t}\bra{t+1} + \ket{t+1}\bra{t} 
			\big).
	\label{Hwalk}
\end{eqnarray}
This is the Hamiltonian of a (continuous-time) quantum walk on a line of length $T+1$. Therefore, $H_{20}$ induces a quantum walk on the ``line'' of states $\ket{\psi_t}$ of the qudit chain of length $L$.


\subsection{Required Evolution Time Analysis}
\label{d20time}

The final step of our model of computation after initializing the qudit chain in the state $\ket{\psi_0}$ and evolving with $H_{20}$ for time $\tau$ is to read out the output of the computation. 
As in Section \ref{d10time}, we need to ensure that the probability of finding the chain of qudits in a state where the computation was performed completely is high. To raise this probability, we choose to pad the program ($K$ sequences of gates) with another $5K$ sequences of identity gates and redo the construction in the previous section. The length of the qudit chain thus becomes $L = (2(6K)-1)(N+1)+2$. The states $\kets{\psi_{t>T/6}}$ (with $T$ modified) now all contain the result of the quantum circuit $U$ in the readout qubit $w_N$, as the relevant gates have been applied to the work qubits in those states. Note that as the extra identity gates pass by, the state of the work qubits does not change. 

The readout procedure consists of two steps. First, measure the qudit $p_{L-K(N+1)}$ in the program register (the qudit with distance from the right end of the chain equal to the length of the original program).
Let us call $p_{20}$ the probability to measure $\bul$ (which would mean the program has moved to the left of the qudit we just measured). When this happens, we are assured we have a state in which the computation was done. Second, we measure $w_N$, the last of the work qubits, and read out the result of the computation $U$.
We will now prove that when we choose to measure $p_{L-K(N+1)}$
at a random time $0\leq \tau\leq \tau_{20}$ with $\tau_{20} = poly(N)$, the probability $p_{20}$ of obtaining the state $\bul$ is close to $\frac{5}{6}$.

To simplify the notation, let us label the states $\ket{\psi_t}$ as $\ket{t}$. In this basis, the Hamiltonian \eqref{ourH} is the negative of the adjacency matrix of a line graph with $T+1$ nodes. For the analysis of time evolution with $H$ we refer the reader to Appendix \ref{d20proof}. We now use the following lemma about a quantum walk on a line (proved in Appendix \ref{d20proof}):
\begin{lem2}
Consider a continuous time quantum walk on a line of length $T+1$, where the Hamiltonian is the negative of the adjacency matrix for the line. Let the system evolve for a time $\tau$ chosen uniformly at random between $0$ and $\tau_{20}$, starting in a position basis state $\ket{c}$.
The probability to measure a state $\ket{t}$ with $t>T/6$ is then $p_{20}\geq \frac{5}{6}-O\left(\frac{T+1}{\tau_{20}}\right)$. 
\end{lem2}

This implies that when we initialize the qudit chain in the state $\ket{\psi_0}$ (corresponding to the leftmost state on the line $\ket{c}=\ket{1}$) and let it evolve with $H$ for a random time $\tau\leq \tau_{20}$ with $\tau_{20} = O(T\log T)$, the probability to find a state with $t>T/6$ is close to $\frac{5}{6}$. Therefore, when we measure the program qudit $p_{L-K(N+1)}$, we will obtain $\bul$ with probability close to $\frac{5}{6}$. 
Finally, when we subsequently measure the work qubit $w_N$, we will obtain the result of the quantum circuit $U$. 

Note that we can also avoid this postselection procedure and simply measure the output qubit. The analysis of the outcome would then follow what we did above in Section \ref{d10time}, resulting in \eqref{sigmayesno} again, with $M$ replaced by $T$.


\section*{Acknowledgments}
The authors would like to thank Dominik Janzing and Jeffrey Goldstone for helpful discussions.
P.W. would like to thank Eddie Farhi's group for their hospitality and the W. M. Keck Foundation for partial support.
D.N. gratefully acknowledges the support from the W. M. Keck Foundation Center for Extreme Quantum Information Theory, and
European Project QAP 2004-IST-FETPI-15848.
P.W. gratefully acknowledges the support by the NSF grants CCF-0726771 and CCF-0746600.


\appendix


\section{Quantum Walk on a Line}
\label{d20proof}

Here we analyze the quantum walk on a line and prove two useful lemmas used in Section \ref{d20time} and Appendix \ref{d10proof}.

Consider a continuous time quantum walk on a line of length $L$, where the Hamiltonian is the negative of the adjacency matrix for the line 
\begin{eqnarray}
	H_1 = - \sum_{j=1}^{L-1} 
	\left(\ket{j}\bra{j+1} + \ket{j+1}\bra{j}\right).
	\label{H1hamiltonian}
\end{eqnarray}
The eigenvalues of this Hamiltonian are
\begin{eqnarray}
	\lambda_j = - 2 \cos \left(\frac{j\pi}{L+1}\right),
	\label{eigenval}
\end{eqnarray}
for $j=1\dots L$, while the corresponding eigenvectors $\ket{\phi^{(j)}} = \sum_{k=1}^{L} \phi^{(j)}_k \ket{k}$ have components
\begin{eqnarray}
	\phi^{(j)}_k = \sqrt{\frac{2}{L+1}} \, 
		\sin \left( \frac{j k \pi}{L+1} \right).
	\label{eigenvec}
\end{eqnarray}
Consider the time evolution of a particular basis state $\ket{c}$. 
The probability of finding the system in a basis state $\ket{m}$ at some time $\tau$ can be found by expanding $\ket{c}$ and $\ket{m}$ in the basis of the eigenvectors \eqref{eigenvec}:
\begin{eqnarray}
	p_{\tau}(m|c) = \left| 
				\bra{m} e^{-iH\tau} \ket{c}
			\right|^2
		= \sum_{j,k=1}^{L} e^{-i (\lambda_j -\lambda_k) \tau} 
			\phi^{(j)}_m
			\phi^{(j)*}_c
			\phi^{(k)*}_m
			\phi^{(k)}_c.
\end{eqnarray}
Because the time evolution (according to the Schr\"odinger equation) is unitary, this probability $p_{\tau}(m|c)$ does not converge. On the other hand, let us define the time average of $p_{\tau}(m|c)$
for time $0\leq\tau\leq\tau_{20}$ as
\begin{eqnarray}
	\bar{p}_{\tau_{20}} (m|c) 
		= \frac{1}{\tau_{20}} \int_0^{\tau_{20}} p_{\tau} (m|c)\,d\tau.
	\label{Paverage}
\end{eqnarray}
As we will show below in Lemma 1, this average probability distribution does converge to a limiting distribution $\pi(m|c)$, defined
as the $\tau_{20}\rightarrow\infty$ limit of the average probability distribution \eqref{Paverage}. All the eigenvalues \eqref{eigenval} are different, so we can express the limiting distribution as
\begin{eqnarray}
	\pi(m|c) 
	= \lim_{\tau_{20}\rightarrow\infty} \bar{p}_{\tau_{20}}(m|c) 
	= \sum_{j=1}^{L} \big|\phi^{(j)}_m\big|^2 
		\big|\phi^{(j)}_c\big|^2,
\end{eqnarray}
which in our case is
\begin{eqnarray}
	\pi(m|c) 
	 = \frac{2+\delta_{m,c} + \delta_{m,L+1-c}}{2(L+1)}.
	 \label{Plimiting}
\end{eqnarray}

According to the following lemma, the average probability \eqref{Paverage} converges to the limiting distribution $\pi(m|c)$.
\begin{lem1}
Consider a continuous time quantum walk on a line of length $L$, where the Hamiltonian is the negative of the adjacency matrix for the line. Let the system evolve for time $\tau \leq \tau_{20}$ chosen uniformly at random, starting in a position basis state $\ket{c}$.
The average probability distribution $\bar{p}_{\tau_{20}}(\cdot|c)$ converges to the limiting probability distribution $\pi(\cdot|c)$ as
\begin{eqnarray}
	\sum_{m=1}^{L} \left| \bar{p}_{\tau_{20}}(m|c) - \pi(m|c) \right|
	\leq 
		O\left(\frac{L}{\tau_{20}}\right).
	\label{limitinglemma}
\end{eqnarray}
\end{lem1}

\begin{proof}
To prove our Lemma 1, we recall Lemma 4.3 of \cite{AAKVwalk:01} for the total variation distance of the probability distribution $\bar{p}_{\tau_{20}}$ from the limiting distribution, saying
\begin{eqnarray}
	\sum_m \left| \bar{p}_{\tau_{20}}(m|c) - \pi(m|c) \right| \leq 
		\frac{2}{\tau_{20}} \sum_{j\neq k} 
		\frac{\big|\phi^{(j)}_c\big|^2}{ |\lambda_j-\lambda_k|}.
	\label{convergebound}
\end{eqnarray}
Using \eqref{eigenval} and \eqref{eigenvec}, we can bound the expression on the right of \eqref{convergebound}. When $j$ is close to $k$, i.e. $|j-k|\leq C_1$, we can obtain
\begin{eqnarray}
	\frac{\big|\phi^{(j)}_c\big|^2}{|\lambda_j-\lambda_k|} < 2.
\end{eqnarray}
On the other hand, for $|j-k|>C_1$ we can write
\begin{eqnarray}
	\frac{\big|\phi^{(j)}_c\big|^2}{|\lambda_j-\lambda_k|} < \frac{C_2}{L+1},
\end{eqnarray}
with $C_1$ and $C_2$ constants independent of $L$. 
Inserting into \eqref{convergebound}, we have
\begin{eqnarray}
	\sum_{m=1}^{L} \left| \bar{p}_{\tau_{20}}(m|c) - \pi(m|c) \right|
	\leq 
		\frac{8 C_1 L}{\tau_{20}} + \frac{C_2 L}{\tau_{20}}
		= O\left(\frac{L}{\tau_{20}}\right),
\end{eqnarray}
which concludes the proof.
\end{proof}

Using Lemma 1, we will now prove a useful result utilized in the time analysis of the $d=20$ HQCA in Section \eqref{d20section}. 

\begin{lem2}
Consider a continuous time quantum walk on a line of length $L$, where the Hamiltonian is the negative of the adjacency matrix for the line. Let the system evolve for a time $\tau \leq \tau_{20}$ chosen uniformly at random, starting in a position basis state $\ket{c}$.
The probability to measure a state $\ket{t}$ with $t>L/6$ is then bounded from below as $p_{20}\geq\frac{5}{6}-O\left(\frac{L}{\tau_{20}}\right)$. 
\end{lem2}

\begin{proof}
The probability to measure a state $\ket{t}$ with $t>L/6$ at time $\tau\leq \tau_{20}$ chosen uniformly at random is
\begin{eqnarray}
	p_{20} = \sum_{m>\frac{L}{6}} 
			 \bar{p}_{\tau_{20}}(m|c).
\end{eqnarray}
Starting with \eqref{limitinglemma}, we have
\begin{eqnarray}
	O\left(\frac{L}{\tau_{20}}\right) 
	&\geq& 
		\sum_{m=1}^{L} 
			\left| \bar{p}_{\tau_{20}}(m|c) - \pi(m|c) \right| \\
	&\geq& 
		\sum_{m>\frac{L}{6}} 
			\left| \bar{p}_{\tau_{20}}(m|c) - \pi(m|c) \right| \\
	&\geq& 
		\left| \sum_{m=\frac{5L}{6}}^{L} 
			 \bar{p}_{\tau_{20}}(m|c) 
		-
		\sum_{m>\frac{L}{6}} 
				\pi(m|c) \right| \\
	&=& 
		\left| p_{20} - \frac{5}{6} + O\left(\frac{1}{L}\right) 
			\right|.
\end{eqnarray} 
Therefore, the probability of finding the chain in state $\ket{\psi_{t>L/6}}$ at a random time $\tau \leq \tau_{20}$ is thus bounded from below by
\begin{eqnarray}
p_{20} \geq \frac{5}{6} - O\left(\frac{L}{\tau_{20}}\right).
\end{eqnarray}
\end{proof}


\section{Diffusion of Fermions on a Line}
\label{d10proof}
We now prove Lemma 3, a result about the mixing of a discrete free fermion gas.

\begin{lem3}
	Consider the state 
\begin{eqnarray}
	\ket{\Psi_0} = b^{\dagger}_{fM+1} b^{\dagger}_{fM+2} \dots b^{\dagger}_{fM+2M} 
	\ket{0}.
\end{eqnarray}
of $2M$ fermions on the right end of a line with $L=(f+2)M$ sites. Let the system evolve for a time chosen uniformly at random between $0$ and $\tau_{10}$ with the Hamiltonian 
\begin{eqnarray}
	H_f = - \sum_{j=1}^{L-1} b_j^\dagger b_{j+1} + h.c.
\end{eqnarray}
and measure the number of fermions in the region $1\leq x \leq fM$. The probability to measure a number greater than $M$
is $p_{10} \geq \frac{f-2}{f+2} - O\left(\frac{L}{\tau_{10}}\right)$.
\end{lem3}

\begin{proof}
Let us start with the outline of the proof. We look at the fermionic system in both first and second quantization to obtain an expression for the time evolution of the creation and annihilation operators in the Heisenberg picture, mapping it to a quantum walk on a line. We then consider the observable $X$, the number of particles sufficiently far from the right end of the line. We will show that when we choose the time to measure $X$ uniformly at random between $0$ and $\tau_{10}$, the expected value we will obtain is approaching a number close to $2M$. To show this, we will express the expected value of $X$ in the time-averaged state of the system using the results from a quantum walk on a line. Finally, because the number of particles in the system is $2M$, we will deduce that the probability to measure a number less than $M$ is then small.

Observe that $H_f$ is the Hamiltonian of a free fermion gas on a line in second quantization (a special case of the XY model). 
The time evolution of the state $\ket{\Psi_0}$ can be obtained by looking at the problem back in the first quantization, where we write $\ket{\Psi_0}$ as
\begin{eqnarray}
	\ket{\Psi_0} = \Big[ \ket{\phi_{fM+1}}\otimes \ket{\phi_{fM+2}} 
		\otimes \cdots \otimes \ket{\phi_{fM+2M}} \Big]^{-},
\end{eqnarray}
with $\ket{\phi_j}=\ket{j}$ in the position basis and $[\,\cdot\,]^{-}$ the standard antisymmetrization operator. We first solve for the time evolution of the corresponding one-particle wavefunction $\ket{\phi_j(\tau)}$ with the Hamiltonian
\begin{eqnarray}
	H_1 = - \sum_{j=1}^{L-1} 
	\left(\ket{j}\bra{j+1} + \ket{j+1}\bra{j}\right), 
	\label{H1hamiltonianAppendix}
\end{eqnarray}
and then obtain the solutions for the many-particle problem by antisymmetrization as
\begin{eqnarray}
	\ket{\Psi(\tau)} = \Big[ \ket{\phi_{fM+1}(\tau)}\otimes 
		\ket{\phi_{fM+2}(\tau)}
		\otimes \cdots \otimes \ket{\phi_{fM+2M}(\tau)} \Big]^{-}.
\end{eqnarray} 
The eigenfunctions of $H_1$ (quantum walk on a line) are plain waves (as in \eqref{eigenval} and \eqref{eigenvec}), and the time evolved states $\ket{\phi_j(\tau)}$ thus readily available. Let us define the unitary matrix $u(\tau)$ by 
\begin{eqnarray}
	\ket{j(\tau)} = \sum_{k=1}^{L} u_{jk}(\tau) \ket{k}.
\end{eqnarray}
Returning to the second quantized system, the time evolution of the creation and annihilation operators in the Heisenberg picture is then
\begin{eqnarray}
	b^{\dagger}_j(\tau) &=& \sum_{k=1}^{L} 
			u_{jk}(\tau) b^{\dagger}_k, \nonumber\\
	b_j(\tau) &=& \sum_{k=1}^{L} 
		u^*_{jk}(\tau) b_k. 
		\label{uheisenberg}
\end{eqnarray}

Consider now the observable $X$, the number of particles in the first $fM$ sites of the line with length $L=(f+2)M$
\begin{eqnarray}
	X &=& \sum_{m=1}^{fM} \hat{n}_{m} \label{Xdef}.
\end{eqnarray}
Its expectation value at time $\tau$ is
\begin{eqnarray}
	E_{\tau}(X) &=& 
	\sum_{m=1}^{fM} \bra{\Psi(\tau)} \hat{n}_{m} \ket{\Psi(\tau)}. 
	\label{Xexpect}
\end{eqnarray}
The number operator for site $m$ is $\hat{n}_m = b_m^{\dagger} b_m$. We can go to the Heisenberg picture and use \eqref{uheisenberg} to write
\begin{eqnarray}
	\bra{\Psi(\tau)} \hat{n}_{m} \ket{\Psi(\tau)}
	&=& 
	\bra{\Psi_0} 
		 b_m^{\dagger}(\tau) b_m(\tau)
	\ket{\Psi_{0}} \\
	&=& 
	\sum_{c=1}^{L} \sum_{d=1}^{L} u_{mc}(\tau)u^*_{md}(\tau) \bra{\Psi_0} 
		  b_c^{\dagger} b_d
	\ket{\Psi_{0}} \\
&=& 
	\sum_{c=1}^{L} 
	 \left| u_{mc}(\tau) \right|^2 \bra{\Psi_0} 
		  b_c^{\dagger} b_c
	\ket{\Psi_{0}} \\
&=& 
	\sum_{c=fM+1}^{L} \underbrace{\left| u_{mc}(\tau) \right|^2
	}_{p_{\tau}(m|c)},
	\label{nexpect10A}
\end{eqnarray}
where each term $\left| u_{mc}(\tau) \right|^2 = p_{\tau}(m|c)$ can be thought of as the probability of finding a particle at site $m$ at time $\tau$ when it started from the site $c$ and performed a quantum walk on a line, according to \eqref{H1hamiltonianAppendix}. Inserting this into \eqref{Xexpect}, the expected number of particles not in the rightmost part of the chain at time $\tau$ is 
\begin{eqnarray}
	E_{\tau}(X) 
	=
	\sum_{c=fM+1}^{L} 
	\left( \sum_{m=1}^{fM} p_{\tau}(m|c) \right)
	.
	\label{nexpect10}
\end{eqnarray}

Let us now choose the time $\tau$ uniformly at random between $0$ and $\tau_{10}$. The average value of $X$ (the expectation value in the time-average state) is
\begin{eqnarray}
	\bar{E}_{\tau_{10}}(X) = \tav{E_{\tau}(X)}.
\end{eqnarray}
For a quantum walk on a line, the time-averaged probability \eqref{Paverage} of finding a particle that started at position $c$ at final position $m$ converges to the limiting distribution \eqref{Plimiting}
according to Lemma 1 \eqref{limitinglemma} proved in Appendix \ref{d20proof}.
Using this fact, we can show that the expectation value $\bar{E}_{\tau_{10}}(X)$ in the time-averaged state converges to the limiting expectation value 
\begin{eqnarray}
 	\bar{E}(X) = \sum_{m\leq fM} \sum_{c>fM} \pi(m|c)
\end{eqnarray}
as
\begin{eqnarray}
	\left|
		\bar{E}_{\tau_{10}}(X) 
		- 
	    \bar{E}(X) \right|
	\leq O\left(\frac{LM}{\tau_{10}}\right).
	\label{expectconverge}
\end{eqnarray}
Recalling the limiting probability distribution for a quantum walk on a line of length $L$ \eqref{Plimiting}, we have
\begin{eqnarray}
 	\bar{E}(X) 
	&=& \sum_{m\leq fM} \sum_{c>fM} \pi(m|c) \\
	&=& fM\times 2M \times 	\frac{2}{2(L+1)}
	+ 2M \times 	\frac{1}{2(L+1)} \\
	&=& 
	2M\left(\frac{f}{f+2}\right) + O\left(1\right)
	.
\end{eqnarray}
Putting this into \eqref{expectconverge}, the average value of $X$ when the time $\tau\leq \tau_{10}$ is chosen uniformly at random is bounded from below as
\begin{eqnarray}
 	\bar{E}_{\tau_{10}}(X) 
	\geq
	 2M\left(\frac{f}{f+2}\right) - O\left(\frac{LM}{\tau_{10}}\right)
	.
\end{eqnarray}
We want to find the probability of measuring $X>M$. 
First, the maximum possible value we could measure at any time is $2M$, the number of particles in the system. Second, the average value $\bar{E}_{\tau_{10}}(X)$ at time $\tau$ chosen randomly is close to $2M$. Therefore, the fraction $\Delta$ of times at which we measure a number significantly lower than $2M$ must be small. Let us bound $\Delta$ in the worst case scenario. This is when each unsuccessful measurement yields $X=M$, and each successful measurement gives us $2M$. 
We then have
\begin{eqnarray}
	\Delta M + (1-\Delta)  2M &\geq& \bar{E}_{\tau_{10}}(X), \\
	\Delta  &\leq& \frac{ 
		 2\bar{E}_{\tau_{10}}(X) - M}{M}.
\end{eqnarray}
Hence we arrive at the desired bound on the probability to measure $X>M$:
\begin{eqnarray}
	p_{10} &=& 1-\Delta 
	\geq \frac{ 
		 2M\left(\frac{f}{f+2}\right) - O\left(\frac{LM}{\tau_{10}}\right)
		  - M
		 }{
		 M
		 }
	= \frac{f-2}{f+2} - O\left(\frac{L}{\tau_{10}}\right).
\end{eqnarray}
\end{proof}




\end{document}